\documentclass[conference]{IEEEtran}

\usepackage{amsmath,amssymb,amsthm}
\usepackage{bm}
\usepackage{graphicx}
\let\labelindent\relax
\usepackage{enumitem}
\usepackage[hidelinks]{hyperref}
\usepackage[capitalize,noabbrev]{cleveref}
\usepackage{microtype}

\theoremstyle{definition}
\newtheorem{remark}{Remark}

\newcommand{\R}{\mathbb{R}}

\newcommand{\E}{\mathbb{E}}
\newcommand{\herm}{\mathsf{H}}
\newcommand{\trans}{\mathsf{T}}
\newcommand{\CN}{\mathcal{CN}}
\newcommand{\Pa}{\mathrm{Pa}}
\DeclareMathOperator{\tr}{tr}
\DeclareMathOperator{\diag}{diag}
\DeclareMathOperator{\rank}{rank}
\DeclareMathOperator{\Cov}{Cov}
\newcommand{\MI}{I}
\newcommand{\param}{\bm{\eta}}
\newcommand{\Ib}{\bm{I}}
\newcommand{\Obs}{\mathcal{O}}

\title{A Differentiable Covariance Calculus for\\
Linear Gaussian Bayesian Networks}

\author{%
\IEEEauthorblockN{Tadashi~Wadayama}\\
\IEEEauthorblockA{Department of Computer Science,
Nagoya Institute of Technology, Nagoya 466-8555, Japan\\
Email: wadayama@nitech.ac.jp}%
}

\begin{document}
\maketitle

\begin{abstract}
Linear Gaussian Bayesian networks, equivalently linear Gaussian structural
equation models, recur across statistics, control, and communications; in the
vector-valued setting that motivates this work, their nodes are vectors and their
edges are matrices. Every quantity of interest is a function of sub-blocks of the
joint covariance, which is itself a classical, differentiable map (the
K-recursion) from the local edge and innovation parameters. Yet the resulting
inference and estimation tasks are usually derived and implemented separately,
per task and per topology. Taking this covariance chart as a single backend, we
build on it a unified, differentiable covariance calculus in which each task
reduces to a few linear-algebra primitives on the one covariance, and automatic
differentiation returns every gradient in a single backward sweep, over
arbitrary vector-valued directed acyclic graphs and parametrizations, including
tied and structured ones. The calculus covers conditioning,
conditional-independence testing through mutual information, maximum-likelihood
estimation with hidden nodes, and the Slepian--Bangs Fisher information with
the local identifiability and Cram\'er--Rao reliability it induces. It is
validated on a linear Gaussian state-space model and a skip-connected
(non-chain) extension against the Kalman recursions, d-separation, and the
Cram\'er--Rao bound.
\end{abstract}


\section{Introduction}
\label{sec:intro}

Linear Gaussian models on directed acyclic graphs (DAGs) recur across
statistics, control, and communications. They are linear structural
equation models (SEMs) and Gaussian graphical models in
statistics~\cite{bollen1989,koller2009}, linear state-space models in
control and signal processing, and cascaded linear channels with
additive Gaussian noise in communications, as in MIMO transceivers,
multi-hop amplify-and-forward relays, and cooperative sensor arrays.
In each case a node carries a Gaussian variable, an edge applies a
linear map, and an independent Gaussian innovation enters at every
non-root node.

In the applications that motivate this work, the nodes are \emph{vectors}
and the edges are \emph{matrices}: an antenna array, a latent feature
vector, or a dynamical state at each node, and a MIMO channel or
processing matrix on each edge. The joint distribution of such a network
is zero-mean Gaussian and is therefore determined entirely by its
covariance, and essentially every quantity of interest (marginals and
conditionals, mutual information, likelihoods, and Fisher information) is a
function of sub-blocks of that
covariance. The covariance is, in turn, a function of the \emph{local}
conditional parameters: the edge matrices and the innovation covariances.
Mapping the local parameters to the global covariance is thus the
computational hub on which all downstream analysis rests.

This local-to-global covariance map is classical; its evaluation as a
differentiable, inverse-free forward operator, together with its relation to
the classical Gaussian-network, path-analysis, and state-space literature,
is developed in the companion paper~\cite{wadayama2026dag}. We take that
covariance backend as given: the object of this paper is the inference and
estimation framework built on it, not the covariance recursion itself.

What is missing is not another formula but an organizing one. Taken
individually, each downstream operation (conditioning by a Schur
complement, a log-determinant mutual information, a node-wise regression
for maximum likelihood) is elementary; assembling all of them, over
arbitrary vector-valued DAGs and in a form that differentiates end to
end, is not. This paper provides that organization: it treats the
local-to-global covariance map as a differentiable \emph{chart} and expresses
inference, likelihood-based estimation, and Fisher-based identifiability as
compositions of a few block-matrix primitives with it. Concretely, we adopt
a single representation of the covariance
map, the \emph{K-recursion}, that
(i)~is \emph{differentiable}: built from matrix products, sums, and
transposes only, it is a smooth computation graph, so
reverse-mode automatic differentiation (AD)~\cite{paszke2019pytorch}
returns the gradient of any downstream scalar with respect to every edge
parameter in a single backward sweep, with no per-topology gradient
derivation;
(ii)~is \emph{vector/matrix-valued}, treating MIMO nodes and matrix edges
natively rather than vectorizing them away; and
(iii)~returns \emph{all node-pair covariance blocks} (including the
non-adjacent cross-covariances that branching and merging topologies
require at a shared descendant) in one inverse-free, topological forward
pass.

On this single operator we build a \emph{differentiable covariance
calculus} for linear Gaussian Bayesian networks: a unified inference and
estimation framework in which every query is a composition of a few
block operations with the chart. It is summarized in the following
contributions.
\begin{itemize}[leftmargin=1.4em]
\item \textbf{A covariance-chart view of Gaussian BN inference.} We use the
differentiable covariance backend of~\cite{wadayama2026dag} as a
\emph{covariance chart} (an explicit, differentiable map from the local
parameters to the covariance family a fixed DAG realizes) and make it the
organizing object for inference, estimation, and identifiability
(\cref{sec:krecursion}).
\item \textbf{A unified calculus on one backend.} Marginalization and
conditioning; mutual information, conditional mutual information, and
conditional-independence testing; maximum-likelihood estimation for fully and
partially observed networks, including hidden nodes;
and the Slepian--Bangs Fisher information, with the local-identifiability
test and Cram\'er--Rao reliability it induces, each of which reduces to
differentiating a scalar through the same operator, so all are obtained from one
backend (\crefrange{sec:inference}{sec:identifiability}).
\item \textbf{Validation against classical references.} On a linear Gaussian
state-space model and a skip-connected (non-chain) extension whose merging
nodes defeat first-order recursions, the calculus is validated end to end
against the Kalman filter and smoother, d-separation, a companion-form
covariance recursion, and the Cram\'er--Rao bound (\cref{sec:examples}).
\end{itemize}
The calculus is realized in an open-source reference implementation,
\texttt{gaussian-bn}, publicly available at
\url{https://github.com/wadayama/gaussian-bn}; every number and figure of
\cref{sec:examples} is produced by it.

\emph{Relation to companion papers.}
This paper is complementary to~\cite{wadayama2026dag,wadayama2026cmi}: those
papers \emph{optimize} information objectives (mutual information and its
multi-terminal conditional variants) through the same backend, whereas the
present paper develops its inference, estimation, and identifiability side;
mutual information appears here only as an information and
conditional-independence measure, not as an optimization objective.

\emph{Organization.}
\cref{sec:prelim} fixes notation, states the model, briefly reviews the
K-recursion, and positions the work within Gaussian-network inference.
\crefrange{sec:inference}{sec:identifiability} build inference, estimation,
identifiability, and reliability on the covariance chart.
\cref{sec:examples} validates the calculus on a linear Gaussian state-space
model and its skip-connected extension, and \cref{sec:conclusion} concludes.

\section{Preliminaries}
\label{sec:prelim}
This section sets the stage for the calculus. \Cref{sec:notation} fixes
notation, \cref{sec:model} states the vector-valued linear Gaussian Bayesian
network model and its (possibly tied and structured) parametrization,
\cref{sec:krecursion} reviews the K-recursion that maps the local parameters
to the covariance chart on which everything else is built, and
\cref{sec:related} places the framework in the Gaussian-network literature.

\subsection{Notation}
\label{sec:notation}
Uppercase italic letters (e.g., $X,V_j$) denote random vectors, and
boldface letters (e.g., $\bm{A},\bm{\Sigma},\bm{K}$) denote deterministic
matrices and vectors. The transpose of $\bm{A}$ is $\bm{A}^{\trans}$. We write
$\bm{\Sigma}\succ\bm{0}$ for symmetric positive definite and
$\bm{\Sigma}\succeq\bm{0}$ for positive semidefinite, and $\bm{A}\succeq\bm{B}$
when $\bm{A}-\bm{B}$ is positive semidefinite (the Loewner order);
$\bm{I}_d$ is the
$d\times d$ identity matrix; $\tr(\cdot)$, $\det(\cdot)$, $\rank(\cdot)$,
and $\E[\cdot]$ are trace, determinant, rank, and expectation; and $\log$
is the natural logarithm.
For a directed acyclic graph (DAG)
$\mathcal{G}=(\mathcal{V},\mathcal{E})$ on the node set
$\mathcal{V}=\{V_1,\dots,V_M\}$,
$\Pa(j)=\{i:(V_i\!\to\!V_j)\in\mathcal{E}\}$ denotes the parent index set
of node $j$. For an index set $A\subseteq\{1,\dots,M\}$, $V_A$ is the
subvector formed by stacking $\{V_j\}_{j\in A}$, and $\bm{K}_{AB}$ is the
block of a covariance matrix indexed by the row set $A$ and column set
$B$; $\Obs\subseteq\{1,\dots,M\}$ denotes the set of observed nodes.

\subsection{Linear Gaussian Bayesian Network Model}
\label{sec:model}
The model is that of the companion paper~\cite{wadayama2026dag}, which states
it over complex vectors for its wireless applications; we state it over the
reals, the convention of the statistical literature, and nothing essential
changes: \cref{rem:complex} records the dictionary between the two cases.
We consider a \emph{linear Gaussian Bayesian network} on a DAG
$\mathcal{G}$ whose nodes are topologically ordered, meaning $i<j$ for
every edge $(V_i\!\to\!V_j)\in\mathcal{E}$, equivalently
$\Pa(j)\subseteq\{1,\dots,j-1\}$. Each node is a Gaussian random vector
$V_j\in\R^{d_j}$ governed by the
linear structural equation
\begin{equation}
V_j=\sum_{i\in\Pa(j)}\bm{A}_{ji}V_i+Z_j,\qquad
Z_j\sim\mathcal{N}(\bm{0},\bm{\Sigma}_j),
\label{eq:sem}
\end{equation}
where $\bm{A}_{ji}\in\R^{d_j\times d_i}$ is the edge transform from
parent $i$ to node $j$, $\bm{\Sigma}_j\succeq\bm{0}$ is the innovation
(conditional) covariance, and the innovations $\{Z_j\}_{j=1}^{M}$ are
mutually independent. A \emph{root} is a parentless node, for
which \eqref{eq:sem} reduces to
$V_r=Z_r\sim\mathcal{N}(\bm{0},\bm{\Sigma}_r)$; we allow any number of
roots. The model is described by the local conditional parameters
$\{\bm{A}_{ji}\}$ and $\{\bm{\Sigma}_j\}$, which need not all be free.
Following the edge factorization of the companion
paper~\cite{wadayama2026dag}, each edge matrix is a product of factors,
\begin{equation}
\bm{A}_{ji}=\bm{A}_{ji}^{(1)}\bm{A}_{ji}^{(2)}\cdots\bm{A}_{ji}^{(L_{ji})},
\label{eq:edgefac}
\end{equation}
each of which is either \emph{learnable} (for example a controllable precoder,
relay gain, or phase profile) or fixed (a known channel), and factors may be
shared across edges. We write $\param$ for the collection of learnable
factors, together with any free innovation covariances; it is the target of
estimation (\cref{sec:estimation}). The case in
which every edge matrix and innovation covariance is itself free,
\begin{equation}
\param=\{\bm{A}_{ji}:(V_i\!\to\!V_j)\in\mathcal{E}\}
       \cup\{\bm{\Sigma}_j:j\in\mathcal{V}\},
\label{eq:params}
\end{equation}
is the extreme with all $L_{ji}=1$ and every factor learnable. Because the
K-recursion composes with any such construction, every quantity below is
differentiable in $\param$ regardless of the parametrization.

\begin{remark}[Tied and structured parameters]
\label{rem:tied}
The free parameters $\param$ need not be the edge matrices themselves. Each
$\bm{A}_{ji}$ may be an arbitrary differentiable function of $\param$: a factor
shared across several edges (weight tying, $\bm{A}_{ji}=\bm{H}_{ji}\bm{F}_i$
with a common $\bm{F}_i$), a low-rank edge $\bm{A}=\bm{U}\bm{V}^{\trans}$, a known
factor multiplying a learnable one, or any other structural constraint. Since the
K-recursion is differentiable, automatic differentiation back-propagates through
this construction, so the gradient with respect to a tied parameter is the sum of
its contributions over all edges on which it appears, in a single backward sweep
and with no per-parametrization derivation. Tying lowers the effective parameter
count $q$, and hence the size of the Fisher information
(\cref{sec:identifiability}); imposing a tie is therefore also a way to eliminate
a gauge and restore identifiability.
\end{remark}

Stacking the nodes, $V=(V_1^{\trans},\dots,V_M^{\trans})^{\trans}$ is an
invertible linear image of the stacked independent innovations
$(Z_1^{\trans},\dots,Z_M^{\trans})^{\trans}$, so the nodes are \emph{jointly
Gaussian}: $V\sim\mathcal{N}(\bm{0},\bm{K})$ with
$\bm{K}=\E[VV^{\trans}]$. The DAG and its local parameters are thus one
graphical representation of this single joint distribution, factoring
$\mathcal{N}(\bm{0},\bm{K})$ into the node conditionals~\eqref{eq:sem}, and
every quantity studied in this paper is a function of sub-blocks of
$\bm{K}$. We denote the node-pair covariance blocks by
$\bm{K}_{jk}=\E[V_jV_k^{\trans}]$; \cref{sec:krecursion} constructs them
from $\param$.

Two features distinguish this setting from the scalar Gaussian Bayesian
networks of classical graphical-model theory. First, nodes are
\emph{vectors} and edges are \emph{matrices} ($\bm{A}_{ji}$ need not be
square), as required by MIMO links, sensor arrays, and latent or state
vectors. Second, the innovation covariance is allowed to be merely PSD,
$\bm{\Sigma}_j\succeq\bm{0}$, which admits deterministic (noiseless) nodes
such as exact state updates or other deterministic mechanisms; the measure-theoretic
status of such degenerate nodes is addressed in \cref{sec:psd}.

\begin{remark}[Zero mean without loss of generality]
\label{rem:mean}
Adding a per-node offset $c_j$, so that
$V_j=c_j+\sum_{i\in\Pa(j)}\bm{A}_{ji}V_i+Z_j$, only shifts the means and leaves
the covariance $\bm{K}$ unchanged; the two moments decouple. Every quantity
studied in this paper (mutual information, conditional independence, and the
Fisher metric) is a function of
$\bm{K}$ alone and is therefore mean-invariant. We accordingly adopt the
zero-mean model ($c_j=\bm{0}$) throughout. The affine mean and the value-level
operations that require it (atomic interventions to a fixed value, and
counterfactuals) are left to separate work.
\end{remark}

\begin{remark}[Complex case]
\label{rem:complex}
We state the formulation for real vectors, $V_j\in\R^{d_j}$. It carries over
verbatim to circular (proper) complex Gaussian networks, such as arise in
modeling MIMO channel statistics: replace the transpose $(\cdot)^{\trans}$ by the
conjugate (Hermitian) transpose $(\cdot)^{\herm}$ and $\mathcal{N}$ by the
circular complex Gaussian $\CN$, under which the Hermitian covariance
$\E[\,\cdot\,(\cdot)^{\herm}]$ fully specifies the
distribution~\cite{schreier_scharf2010}. The only
quantitative change is the constant in the information and Fisher measures
(\cref{sec:mi,sec:fisher}), where the factor $\tfrac12$ becomes $1$, reflecting
the real versus circular-complex degrees of freedom.
\end{remark}

\subsection{K-Recursion: A Brief Review}
\label{sec:krecursion}
The node-pair covariance blocks $\bm{K}_{jk}=\E[V_jV_k^{\trans}]$ follow from
the local parameters $\param$ by a single topological forward pass.
Substituting the structural equation~\eqref{eq:sem} and using independence of
the innovations gives, in topological order,
\begin{equation}
\begin{aligned}
\bm{K}_{jk}&=\sum_{i\in\Pa(j)}\bm{A}_{ji}\bm{K}_{ik}
  \qquad (k<j),\\[2pt]
\bm{K}_{jj}&=\sum_{i,i'\in\Pa(j)}\bm{A}_{ji}\bm{K}_{ii'}\bm{A}_{ji'}^{\trans}
  +\bm{\Sigma}_j,
\end{aligned}
\label{eq:krec}
\end{equation}
with $\bm{K}_{rr}=\bm{\Sigma}_r$ at a root $r$ and
$\bm{K}_{kj}=\bm{K}_{jk}^{\trans}$, so only the blocks with $j\geq k$ need be
stored. The self-block carries the \emph{parent cross-covariances}
$\bm{K}_{ii'}$ ($i\neq i'$), which are what make the construction correct at a
node that merges several parents.

Collecting the edge transforms into the strictly block-lower-triangular
matrix $\bm{A}$, whose $(j,i)$ block is $\bm{A}_{ji}$ for $i\in\Pa(j)$ and zero
otherwise, and the innovation covariances into
$\bm{\Sigma}=\diag(\bm{\Sigma}_1,\dots,\bm{\Sigma}_M)$, the
recursion~\eqref{eq:krec} evaluates the closed form
$\bm{K}=(\Ib-\bm{A})^{-1}\bm{\Sigma}(\Ib-\bm{A})^{-\trans}$ by topological
forward substitution, without ever forming the inverse. The two routes agree,
but not in cost: a direct evaluation of the closed form inverts (or solves
with) the full $D\times D$ matrix $\Ib-\bm{A}$, with $D=\sum_j d_j$,
regardless of the graph, whereas the recursion touches only the blocks the
edges make nonzero, spending a few small matrix products per edge, and keeps
every intermediate quantity an explicit covariance block; this is why the
recursion, rather than the closed form, is the backend used throughout.
Since it uses only
matrix products, sums, and transposes, the map
$\Phi_{\mathcal{G}}:\param\mapsto\bm{K}$ is a smooth computation
graph, and reverse-mode AD returns the gradient of any downstream scalar with
respect to every entry of $\param$ in one backward sweep. In other words,
when the forward
recursion~\eqref{eq:krec} is executed in an AD framework, every intermediate
block it forms is recorded, together with the elementary operation that
produced it, as a node of one connected computation graph running from
$\param$ to $\bm{K}$. A scalar built afterwards from blocks of $\bm{K}$
therefore sits at the end of an unbroken chain of operations with known
derivatives, and the backward sweep is nothing more than the chain rule
applied along that chain in reverse, so no gradient formula is ever derived
by hand.

We regard
$\Phi_{\mathcal{G}}$ as a \emph{covariance chart}, an explicit differentiable
parametrization of the covariances a fixed DAG realizes, and build the
inference, estimation, and identifiability framework of the sequel on it. Its
derivation, numerical treatment, and relation to prior covariance
constructions are detailed in the companion paper~\cite{wadayama2026dag}; here
we use it as a given backend.

\begin{remark}[Degenerate noise]
\label{sec:psd}
With $\bm{\Sigma}_j\succeq\bm{0}$ (\cref{sec:model}), \eqref{eq:krec} remains
valid on the whole PSD cone, since it expresses only the bilinearity of
covariance under linear maps; a deterministic node ($\bm{\Sigma}_j=\bm{0}$)
is then the noiseless linear function
$V_j=\sum_{i\in\Pa(j)}\bm{A}_{ji}V_i$ of its parents. Whether such nodes are
admissible depends on the query, not on the model. Every operation on the
covariance itself (the recursion, marginal covariances, forward sampling)
tolerates them; a quantity defined through a density (a likelihood, a Fisher
information, or a mutual information across a deterministic link) instead
requires the covariance block it touches to be positive definite. The
requirement falls on the block of $\bm{K}$, not on $\bm{\Sigma}_j$: a
noiseless node whose parents carry enough noise still has a positive-definite
marginal covariance, and hence a density. We note the condition where it
arises.
\end{remark}

\subsection{Related Work}
\label{sec:related}
The differentiable covariance backend used here (the K-recursion) and its
lineage in the Gaussian-influence-diagram, path-analysis, and state-space
literatures are established in the companion paper~\cite{wadayama2026dag}; we
do not repeat that discussion, and instead position the present framework
within Gaussian-network inference.

\emph{Inference and estimation in Gaussian graphical models.}
Exact inference in Gaussian Bayesian networks and graphical
models (marginals, conditionals, and conditional independence) is
classical~\cite{koller2009,lauritzen1996,bishop2006}, with Gaussian influence diagrams
and recursive belief-network constructions providing early algorithmic
treatments~\cite{shachter1989,geiger1994}. The same conditioning identities
drive Gaussian process regression~\cite{rasmussen2006}, which differs only in
the origin of the covariance: a kernel evaluated on inputs there, the
covariance chart of a parametrized DAG here (\cref{sec:overview}).
Parameter estimation from data,
including the latent-variable case, proceeds classically by maximum likelihood
and the expectation--maximization algorithm~\cite{dempster1977}; here we maximize
the Gaussian likelihood directly by gradient descent through the differentiable
covariance map. The linear structural-equation-model
viewpoint~\cite{bollen1989} underlies that Gaussian likelihood.

\emph{Identifiability and conditional independence.}
Which parameters of a linear Gaussian model are recoverable from a given set
of observed nodes is an identifiability question; trek separation and the
algebraic-statistics view of Gaussian graphical models characterize the
covariance structure and the associated conditional
independences~\cite{sullivant2010,drton2009}. We treat identifiability and
reliability operationally, through the rank and inverse of a Fisher information
pulled back to the parameters~\cite{kay1993} (\cref{sec:identifiability}).
This rank test touches two classical strands: the pulled-back metric is the
Fisher--Rao metric of the Gaussian family in the sense of information
geometry~\cite{amari2000}, and a parameter at which it loses rank is a
\emph{singular} point of the statistical model in the sense of singular
learning theory~\cite{watanabe2009}, hidden-node Gaussian networks being
canonical examples. We use only the local verdict, the rank and null space at
the parameter at hand, and none of the asymptotic machinery of the singular
case.

\emph{This work.}
These operations are usually developed one at a time, and often per
topology. We obtain all of them (inference, estimation, and Fisher-based
identifiability and Cram\'er--Rao reliability) for arbitrary vector-valued
Gaussian Bayesian networks from one differentiable covariance backend, with
gradients supplied by automatic differentiation and no per-topology
derivation. The complementary use of the same backend for
mutual-information \emph{optimization} is developed in the companion
paper~\cite{wadayama2026dag}.

\section{Inference}
\label{sec:inference}
Inference is the first layer of the calculus. \Cref{sec:overview} casts every
probabilistic query about the network as block algebra on the single
covariance $\bm{K}$; \cref{sec:conditioning,sec:mi} record the basic queries,
from marginals and conditionals to mutual information and
conditional-independence tests; and \cref{sec:sampling} adds their generative
counterpart, forward sampling.

\subsection{Overview: Inference as Block Algebra on the Covariance Chart}
\label{sec:overview}
A linear Gaussian Bayesian network is a probabilistic model, and
\emph{inference} means answering probabilistic questions about its nodes:
what is the distribution of some variables of interest, and how does that
distribution change once other variables are observed? Because the joint law
is the zero-mean Gaussian $V\sim\mathcal{N}(\bm{0},\bm{K})$ of
\cref{sec:model}, both questions have closed-form answers that involve only
sub-blocks of the single covariance matrix $\bm{K}$ produced by
the K-recursion.

Let $A$ be a set of \emph{query} nodes, $B$ a set of \emph{observed} nodes,
and $R$ the \emph{remaining} nodes (neither queried nor observed). Before any
observation, the prior (marginal) law of $V_A$ is read directly off the
diagonal block,
\begin{equation}
V_A\sim\mathcal{N}(\bm{0},\bm{K}_{AA}).
\label{eq:ov_marginal}
\end{equation}
After observing $V_B=\bm{b}$, the posterior law of $V_A$ is again Gaussian,
\begin{equation}
V_A\mid(V_B=\bm{b})\ \sim\
\mathcal{N}\!\big(\,\underbrace{\bm{K}_{AB}\bm{K}_{BB}^{-1}\bm{b}}_{\text{mean}},\
\underbrace{\bm{K}_{A\mid B}}_{\text{cov.}}\,\big),
\label{eq:ov_posterior}
\end{equation}
whose covariance is the Schur complement of the observed block,
\begin{equation}
\bm{K}_{A\mid B}=\bm{K}_{AA}-\bm{K}_{AB}\bm{K}_{BB}^{-1}\bm{K}_{BA}.
\label{eq:ov_schur}
\end{equation}
This is Bayesian updating in closed form: observing $B$ pulls the mean toward
the data through the gain $\bm{K}_{AB}\bm{K}_{BB}^{-1}$ and shrinks the
covariance from the prior $\bm{K}_{AA}$ to the posterior
$\bm{K}_{A\mid B}\preceq\bm{K}_{AA}$, the Schur complement of the observed
block $\bm{K}_{BB}$ in the joint covariance. These are precisely the
conditioning identities that underlie Gaussian process (GP)
regression~\cite{rasmussen2006}; the only difference is the origin of the
covariance, which the K-recursion generates from the network parameters
rather than from a kernel evaluated on inputs.

\begin{figure}[t]
\centering
\includegraphics[width=\linewidth]{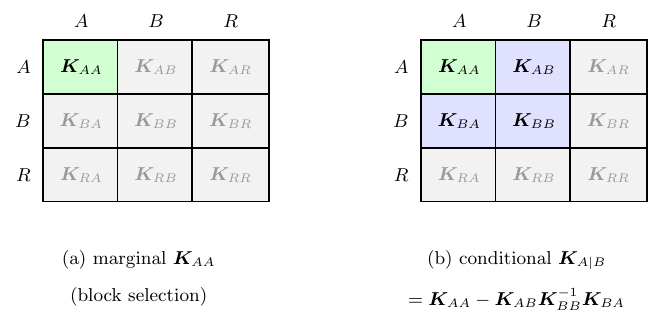}
\caption{Inference reads sub-blocks of the single covariance matrix
$\bm{K}$. (a)~The prior (marginal) covariance of the query nodes $A$
is the diagonal block $\bm{K}_{AA}$ (green); no other block is touched.
(b)~The posterior covariance after observing $B$ is the Schur complement
$\bm{K}_{A\mid B}=\bm{K}_{AA}-\bm{K}_{AB}\bm{K}_{BB}^{-1}\bm{K}_{BA}$, built
from the anchor block $\bm{K}_{AA}$ (green) and the blocks coupling the query
$A$ with the observation $B$, namely $\bm{K}_{AB},\bm{K}_{BA},\bm{K}_{BB}$
(blue). The remaining nodes $R$ (grey), neither queried nor observed, are
marginalized out, i.e.\ their rows and columns are simply dropped.}
\label{fig:block_algebra}
\end{figure}

Because the K-recursion (\cref{sec:krecursion}) has already produced
\emph{every} node-pair block $\bm{K}_{jk}$, the set-indexed blocks used
above are simply gathered from it ($\bm{K}_{AA}$ stacks the blocks
$\bm{K}_{jk}$ with $j,k\in A$, and $\bm{K}_{AB},\bm{K}_{BB}$ likewise), so no
per-query recomputation is needed. \Cref{fig:block_algebra} reads the two
formulas off this assembled matrix.
In panel~(a) the marginal of $V_A$ is the green diagonal block $\bm{K}_{AA}$.
In panel~(b) the posterior covariance $\bm{K}_{A\mid B}$ combines the anchor
block $\bm{K}_{AA}$ (green) with the blocks $\bm{K}_{AB},\bm{K}_{BA}$ that
couple the query to the observation and the observed block $\bm{K}_{BB}$
(blue); the unrelated nodes $R$ (grey) are dropped. Conditioning therefore
\emph{shrinks} $\bm{K}_{AA}$ by exactly the information that the observation
$B$ carries about $A$.

Every inference quantity in this paper is built the same way, from a handful
of operations on sub-blocks of the one matrix $\bm{K}$: \emph{block
selection} (marginals, \cref{eq:ov_marginal}), the \emph{Schur complement}
(conditioning, \cref{eq:ov_schur}), the \emph{log-determinant} of a block
(entropies, mutual information, and Gaussian log-likelihoods,
\cref{sec:mi}), and the \emph{linear solve} $\bm{K}_{BB}^{-1}(\cdot)$
(posterior means and node-wise regressions). We call this \emph{block
algebra} on the covariance chart. The chart supplies the object and the four
primitives the operations; together they constitute the covariance calculus
of the title, and \cref{sec:conditioning,sec:mi} develop it for the basic
queries.

Two properties make this the organizing principle of the paper. First, each
operation is a smooth map (evaluated in practice by one Cholesky
factorization per solve and log-determinant), so composing it with the chart
$\Phi_{\mathcal{G}}:\param\mapsto\bm{K}$ makes any inference quantity
a single differentiable function of the learnable parameters $\param$; one
reverse-mode AD sweep then returns its gradient with respect to every entry of
$\param$, with no query- or topology-specific derivation. Second, the same operations, over the same chart, drive the later
sections: a marginal log-likelihood is a log-determinant plus a trace
(\cref{sec:estimation}); the Fisher information is built from derivatives of
an observed block $\bm{K}_{\Obs\Obs}$ (\cref{sec:identifiability}). A query
that conditions on, or
takes the density of, a node set requires the corresponding block of
$\bm{K}$ to be positive definite (\cref{sec:psd}).

\subsection{Marginalization and Conditioning}
\label{sec:conditioning}
We now record the two basic queries of \cref{sec:overview} in the form used by
the rest of the paper.

\emph{Marginalization.} For a query set $A$, the marginal law is
\begin{equation}
V_A\sim\mathcal{N}(\bm{0},\bm{K}_{AA}),
\label{eq:marginal}
\end{equation}
whose covariance is the diagonal block $\bm{K}_{AA}$, read off the blocks of
\cref{sec:krecursion} with no further computation.

\emph{Conditioning.} Let $B$ be an observed set with $\bm{K}_{BB}\succ\bm{0}$.
Given $V_B=\bm{b}$, the query $V_A$ is Gaussian with
\begin{align}
\E[V_A\mid V_B=\bm{b}] &= \bm{K}_{AB}\bm{K}_{BB}^{-1}\bm{b},
  \label{eq:cond_mean}\\
\Cov[V_A\mid V_B] &= \bm{K}_{A\mid B}
  =\bm{K}_{AA}-\bm{K}_{AB}\bm{K}_{BB}^{-1}\bm{K}_{BA}.
  \label{eq:cond_cov}
\end{align}
The posterior mean~\eqref{eq:cond_mean} is the minimum-mean-square-error
estimate of $V_A$ from the observation $\bm{b}$ and is linear in $\bm{b}$;
although the model is zero-mean, conditioning on a specific value yields this
nonzero data-driven mean. The posterior covariance~\eqref{eq:cond_cov}, by
contrast, does not depend on $\bm{b}$: the observed value relocates the
distribution, but the reduction in uncertainty is fixed by the model alone.

The condition $\bm{K}_{BB}\succ\bm{0}$ is the density-level requirement of
\cref{sec:psd}: it is what makes $\bm{K}_{BB}^{-1}$, and hence the conditional
law, well-defined, and it holds automatically whenever the observed nodes
carry a positive-definite innovation. Numerically we never form
$\bm{K}_{BB}^{-1}$. A Cholesky factor $\bm{K}_{BB}=\bm{L}\bm{L}^{\trans}$ is
computed once; the gain $\bm{K}_{AB}\bm{K}_{BB}^{-1}$ and the
mean~\eqref{eq:cond_mean} follow from triangular solves against $\bm{L}$, and
the Schur complement~\eqref{eq:cond_cov} is re-symmetrized as
$\tfrac12(\bm{K}_{A\mid B}+\bm{K}_{A\mid B}^{\trans})$ to discard the
antisymmetric part introduced by finite-precision arithmetic. The
log-determinants of the information measures in \cref{sec:mi} are evaluated by
the same Cholesky route.

\subsection{Mutual Information and Conditional Independence}
\label{sec:mi}
The information measures follow from the same conditional covariances by a
log-determinant~\cite{coverthomas2006}. For disjoint node sets $A,B,C$, the
mutual information and
the conditional mutual information (CMI) are
\begin{align}
\MI(V_A;V_B) &= \tfrac12\big(\log\det\bm{K}_{AA}-\log\det\bm{K}_{A\mid B}\big),
  \label{eq:mi}\\
\MI(V_A;V_B\mid V_C) &= \tfrac12\big(\log\det\bm{K}_{A\mid C}
  -\log\det\bm{K}_{A\mid BC}\big),
  \label{eq:cmi}
\end{align}
in nats, where each conditional covariance $\bm{K}_{A\mid\cdot}$ is the Schur
complement of \cref{eq:cond_cov}; for circular-complex networks the factor
$\tfrac12$ becomes $1$ (\cref{rem:complex}). Both measures require the
conditioned blocks to be positive definite (\cref{sec:psd}), and both are
evaluated from Cholesky log-determinants of blocks already assembled in
\cref{sec:conditioning}.

Because $\bm{K}$ is a smooth function of $\param$, so are \eqref{eq:mi} and
\eqref{eq:cmi}; the differentiable \emph{optimization} of the mutual
information over the controllable parameters is the subject of the companion
paper~\cite{wadayama2026dag}, with conditional-mutual-information objectives for
multi-terminal networks in~\cite{wadayama2026cmi}. Here the CMI serves instead as a test of
\emph{conditional independence}: for jointly Gaussian variables,
\begin{equation}
V_A\perp V_B\mid V_C \iff \MI(V_A;V_B\mid V_C)=0,
\label{eq:ci}
\end{equation}
which holds exactly when the conditional cross-covariance
$\bm{K}_{AB\mid C}=\bm{K}_{AB}-\bm{K}_{AC}\bm{K}_{CC}^{-1}\bm{K}_{CB}$ is the
zero matrix. Testing conditional independence is therefore a
zero-test on a single Schur-complement block. The test is a basic primitive
in several roles: constraint-based structure learning queries exactly such a
CI oracle, model diagnostics check the independences a hypothesized structure
implies (for instance, whether a first-order Markov model suffices,
\cref{sec:exp-skip}), and a vanishing CMI certifies a sensor as redundant
given those already selected, or an observable as leakage-free given the
public variables.

\emph{Relation to the graph.} Two notions of conditional independence should
be kept apart. The test~\eqref{eq:ci} is \emph{parameter-specific}: it
concerns the particular $\param$ at hand. A \emph{structural} independence,
by contrast, holds for \emph{all} parameter values and is read from the graph
by d-separation~\cite{koller2009}, equivalently by trek
separation~\cite{sullivant2010} in the Gaussian case. A structural
independence implies the numerical test~\eqref{eq:ci} identically in
$\param$; the converse can fail only on a measure-zero set of parameters,
where a CMI vanishes without a graphical reason. In the vector-valued setting
the test is also finer-grained than the graph: d-separation is all-or-nothing
at the node level, whereas a low-rank edge can leave $\bm{K}_{AB\mid C}$
nonzero but rank-deficient, so that only certain linear combinations of $V_A$
are conditionally independent of $V_B$, a sub-block structure the
Schur-complement test resolves and trek separation
characterizes~\cite{sullivant2010}. The K-recursion supplies the
covariance blocks for the numerical test, complementing the graphical
criterion.

\subsection{Forward Sampling}
\label{sec:sampling}
The topological forward pass that propagates second moments in the K-recursion
also \emph{draws} samples from the network. Traversing the nodes in topological
order, we simulate the structural equation~\eqref{eq:sem} directly: at each node
we draw a standard normal $\bm{\xi}_j\sim\mathcal{N}(\bm{0},\bm{I}_{d_j})$ and set
\begin{equation}
V_j=\sum_{i\in\Pa(j)}\bm{A}_{ji}V_i+\bm{L}_j\bm{\xi}_j,
\label{eq:sample}
\end{equation}
where $\bm{L}_j$ is any factor of the innovation covariance,
$\bm{\Sigma}_j=\bm{L}_j\bm{L}_j^{\trans}$, so that $\bm{L}_j\bm{\xi}_j$ realizes
the innovation $Z_j$ of \eqref{eq:sem}. One such pass produces one draw of the
full vector $V\sim\mathcal{N}(\bm{0},\bm{K})$, and $N$ independent runs give $N$
i.i.d.\ samples.
A Cholesky factor serves when $\bm{\Sigma}_j\succ\bm{0}$, and an
eigendecomposition factor when $\bm{\Sigma}_j$ is singular, so that a
deterministic node ($\bm{\Sigma}_j=\bm{0}$) simply propagates its parents. The
pass is linear in the size of the graph and, like the covariance recursion,
forms neither $\bm{K}$ nor any inverse: it realizes the joint law node by node.

Sampling is the generative counterpart of the inference queries above and the
data source for the estimators of \cref{sec:estimation}. The sample covariance
of the draws converges to the analytic $\bm{K}$ of \cref{sec:krecursion}, which
both cross-checks the backend and enables Monte-Carlo evaluation of quantities
that lack a closed form, for example nonlinear statistics of $V$ or empirical
mutual-information and estimator-variance studies. In the complex case the
innovations are circularly symmetric (\cref{rem:complex}).

\section{Estimation}
\label{sec:estimation}
So far the parameters have been given, and the covariance chart has run
\emph{forward}, from $\param$ to $\bm{K}$ to the queries of
\cref{sec:inference}. Estimation is the inverse problem: from data, recover
the learnable parameters $\param$ that generated it. We treat two regimes,
full observation by maximum likelihood (\cref{sec:mle}) and partial
observation with hidden nodes by the marginal likelihood (\cref{sec:partial});
both minimize one differentiable objective through the chart.

\subsection{Full Observation: Maximum Likelihood}
\label{sec:mle}
From $N$ independent realizations $\bm{v}^{(1)},\dots,\bm{v}^{(N)}$ of the
full node vector $V$, form the sample covariance
\begin{equation}
\hat{\bm{K}}=\frac{1}{N}\sum_{n=1}^{N}\bm{v}^{(n)}(\bm{v}^{(n)})^{\trans}.
\label{eq:samplecov}
\end{equation}
Up to an additive constant and a positive factor, the negative log-likelihood is
\begin{equation}
L(\param)=\log\det\bm{K}+\tr\!\big(\bm{K}^{-1}\hat{\bm{K}}\big),
\label{eq:nll}
\end{equation}
a function of the parameters through the covariance chart
$\param\mapsto\bm{K}$; the maximum-likelihood estimate is exactly its
minimizer, $\hat{\param}=\arg\min_{\param}L(\param)$. This objective is
natural on two counts. First, maximum likelihood is consistent and
asymptotically efficient. Second, it is a Kullback--Leibler projection:
minimizing $L$ is the same as minimizing the divergence
$\mathrm{KL}\big(\mathcal{N}(\bm{0},\hat{\bm{K}})\,\|\,\mathcal{N}(\bm{0},\bm{K})\big)$,
so the estimate is the covariance closest to the empirical $\hat{\bm{K}}$
within the family the DAG can realize.

Since that chart is a differentiable computation graph
(\cref{sec:krecursion}), so is $L$: one reverse-mode AD sweep returns the
exact gradient $\nabla_{\param}L$, propagated through both the K-recursion and
the construction of the edges from $\param$, so $L$ is minimized by gradient
descent in a
positive-definite-preserving parametrization, with no per-node or
per-topology gradient derivation. One caveat is in order: $L$ is convex in
the precision $\bm{K}^{-1}$ but not, in general, in $\param$, since the chart
is polynomial. With free per-node parameters the full-observation MLE is
available in closed form by node-wise regression, but under the tied,
structured, or hidden-node parametrizations of interest here, gradient
descent is guaranteed a stationary point only; random initializations
sufficed in \cref{sec:examples}, and the Fisher analysis of
\cref{sec:identifiability} flags the flat gauge directions along which
minimizers form orbits. This is the estimator the framework
provides; unchanged, it handles hidden nodes (\cref{sec:partial}) and any structured
parametrization of the edges (\cref{sec:model}).

A nonzero data mean, when present, is removed by centering before forming
\eqref{eq:samplecov}, which leaves \eqref{eq:nll} unchanged (\cref{rem:mean}).

\subsection{Partial Observation: Marginal Likelihood}
\label{sec:partial}
In many networks not every node can be measured. A hidden common cause, an
unmeasured relay or state node, or a latent factor is seen only through its
effect on the rest of the network. Writing $\Obs$ for the observed nodes, we
must recover $\param$ from data on $\Obs$ alone. The difficulty is
statistical rather than procedural: a hidden node cannot be fitted by its own
regression, and marginalizing it out couples the observed nodes, so no
node-wise closed form survives. On the chart, however, marginalization is
mere block selection (\cref{sec:conditioning}), and the estimator is the one
of \cref{sec:mle} unchanged: minimize the likelihood of the observed
subvector by gradient descent.

Since $V_\Obs\sim\mathcal{N}(\bm{0},\bm{K}_{\Obs\Obs})$, the observed-data
negative log-likelihood is, up to an additive constant and a positive factor,
\begin{equation}
L(\param)=\log\det\bm{K}_{\Obs\Obs}
   +\tr\!\big(\bm{K}_{\Obs\Obs}^{-1}\bm{S}\big),
\label{eq:marglik}
\end{equation}
where $\bm{S}$ is the sample covariance of the observed nodes and
$\bm{K}_{\Obs\Obs}$ is the observed block of $\bm{K}$ from the
K-recursion. This is \eqref{eq:nll} restricted to $\Obs$; with per-sample
missing data it averages the per-pattern terms
$\log\det\bm{K}_{O_nO_n}+\bm{y}_n^{\trans}\bm{K}_{O_nO_n}^{-1}\bm{y}_n$ over the
patterns, where $O_n$ and $\bm{y}_n$ are the observed set and observed values of
sample $n$. Forming $\bm{K}_{\Obs\Obs}$ with its Cholesky log-determinant and
solve costs $O(D_\Obs^3)$ in the observed dimension
$D_\Obs=\sum_{j\in\Obs}d_j$, and one AD sweep through the K-recursion again
returns the exact gradient $\nabla_{\param}L$, for any parametrization
including the tied and structured parameters of \cref{rem:tied}.
This direct route complements expectation--maximization~\cite{dempster1977}
rather than replacing it: where closed-form E- and M-steps exist, EM remains
effective, but every tied or structured parametrization requires its own
M-step derivation, whereas \eqref{eq:marglik} is differentiated through the
covariance chart with no such per-parametrization work.

\section{Identifiability and Reliability}
\label{sec:identifiability}
Estimation raises two further questions: \emph{identifiability}, whether
$\param$ can be recovered from the observed nodes at all or whether different
parameters produce the same observed distribution, and \emph{reliability},
how precisely an identifiable $\param$ can be estimated, that is, how much
the estimate scatters around the truth. Both are
answered by one object, the Fisher information, which measures how sensitively
the observed distribution responds to a change in $\param$; a direction of
$\param$ that barely moves the distribution is hard to estimate, and one that
moves it sharply is easy. \Cref{sec:fisher} constructs this metric on the
covariance chart, and \cref{sec:gauge} reads local identifiability off its
rank and Cram\'er--Rao reliability off its inverse.

\subsection{The Pullback (Slepian--Bangs) Fisher Metric}
\label{sec:fisher}
Because the observed nodes are zero-mean Gaussian,
$V_\Obs\sim\mathcal{N}(\bm{0},\bm{K}_{\Obs\Obs})$, all of that sensitivity is
carried by how the observed covariance moves with $\param$. Identify $\param$
with the real vector $(\eta_1,\dots,\eta_q)^{\trans}$ of its $q$ free scalar
coordinates: by \cref{sec:model} these are exactly the parameters to be
estimated, whereas any fixed, known factors (a known channel, a fixed
structure) enter $\bm{K}_{\Obs\Obs}$ only as constants and are held fixed in the
derivatives below.

Recall the general definitions. For a parametric family of densities
$p_{\param}$, the per-observation Fisher information matrix is the covariance
of the score, $G_{ab}(\param)=\E[s_a s_b]$, where
$s_c=\partial\log p_{\param}(v)/\partial\eta_c$ is the score with respect to
the $c$th coordinate. Information geometry reads
$G$ as a Riemannian metric on the parameter space, the Fisher--Rao
metric~\cite{amari2000}: between neighboring models,
$\mathrm{KL}\big(p_{\param}\,\|\,p_{\param+\bm{\delta}}\big)
=\tfrac12\bm{\delta}^{\trans}G(\param)\bm{\delta}
+o(\lVert\bm{\delta}\rVert^{2})$, so $G$ is exactly the local
distinguishability of parameters from data. For the zero-mean Gaussian family
at hand the parameters enter only through $\bm{K}_{\Obs\Obs}$, and the
expectation evaluates in closed form, giving the per-observation Fisher
metric $G=(G_{ab})\in\R^{q\times q}$ on the parameters, whose $(a,b)$ entry is
\begin{equation}
G_{ab}(\param)=\tfrac12\,\tr\!\Big[
  \bm{K}_{\Obs\Obs}^{-1}\frac{\partial\bm{K}_{\Obs\Obs}}{\partial\eta_a}\,
  \bm{K}_{\Obs\Obs}^{-1}\frac{\partial\bm{K}_{\Obs\Obs}}{\partial\eta_b}\Big],
\label{eq:fisher}
\end{equation}
for $a,b\in\{1,\dots,q\}$: the Slepian--Bangs formula~\cite{kay1993}, the
closed form the general definition takes for a zero-mean Gaussian family (the
factor $\tfrac12$ becomes $1$ in the circular-complex case,
\cref{rem:complex}; \cref{app:sb} gives the derivation). Every ingredient comes
from the K-recursion: $\bm{K}_{\Obs\Obs}$ is a forward pass and its Jacobian
$\partial\bm{K}_{\Obs\Obs}/\partial\eta_a$ follows by automatic
differentiation, one sweep per parameter, so $G$
is obtained \emph{analytically}, with no sampling, in $O(q)$ AD sweeps. Since
the derivative computation can itself be retained as part of the computation
graph, the same construction also supports nested AD: smooth scalar criteria
built from $G$, such as Cram\'er--Rao design objectives based on $G^{-1}$ or a
regularized inverse, can be differentiated again with respect to the underlying
parameters or design variables.

In this sense $G$ is not a new model-specific object but the pullback of the
Fisher--Rao metric of the Gaussian family along the covariance chart,
summarizing exactly the
parameter directions visible through the observed distribution. It is the common
root of what follows (\cref{sec:gauge}): its rank decides identifiability and
its (pseudo-)inverse bounds the estimation error.

\subsection{Local Identifiability, Gauges, and Reliability}
\label{sec:gauge}
The parameters are \emph{locally identifiable} from $\Obs$ exactly when the
Fisher metric has full rank, $\rank G=q$, with $q$ the number of parameters. A
rank deficiency exposes a \emph{gauge}: a null direction $\bm{u}$ with
$G\bm{u}=\bm{0}$ is a change of $\param$ that leaves $\bm{K}_{\Obs\Obs}$, and
hence the observed distribution, unchanged to first order. Motion along such a
direction is invisible in the data, so the parameters it mixes cannot be told
apart; $\ker G$ is the tangent space of these gauge orbits, and its dimension
$q-\rank G$ counts the undetermined directions. Such a point is, in the
terminology of statistical learning theory, a \emph{singular} point of the
model~\cite{watanabe2009}: the map from parameters to distributions fails to
be locally one-to-one, so distinct parameters realize the same observed law,
as the following example shows exactly.

The canonical example is the latent scale gauge. Let a hidden node $H$ of
variance $\sigma^2$ feed two observed nodes, $Y_1=aH$ and $Y_2=bH$. The
observed covariance depends on $(a,b,\sigma^2)$ only through the products
$a^2\sigma^2$, $b^2\sigma^2$, and $ab\,\sigma^2$, so the reparametrization
$(a,b,\sigma^2)\mapsto(a/t,\,b/t,\,t^2\sigma^2)$ leaves $\bm{K}_{\Obs\Obs}$
invariant for every $t\neq0$. Its generator lies in $\ker G$: the latent scale
is unidentifiable, and only gauge-invariant combinations (here $a\sigma$ and
$b\sigma$) can be recovered. Reporting $\rank G$, the smallest eigenvalues of
$G$ (which flag near-degeneracy), and the null eigenvectors mapped back to the
parameters therefore diagnoses exactly which parameters, or which combinations,
the observed nodes leave undetermined. In numerical use this verdict is read
from the spectrum of $G$ rather than by an exact symbolic rank test: exact
zeros mark gauges of the covariance chart, whereas small but nonzero
eigenvalues mark near-gauges, with correspondingly large Cram\'er--Rao standard
errors.

The same metric also quantifies how well the identifiable parameters can be
estimated. Since $G$ is the per-observation Fisher information, the
Cram\'er--Rao bound states that any unbiased estimator $\hat{\param}$ obeys
\begin{equation}
\Cov(\hat{\param})\succeq G^{-1}.
\label{eq:crb}
\end{equation}
The maximum-likelihood estimator of \cref{sec:estimation} attains this bound
asymptotically: for $N$ independent observations its error covariance approaches
$\tfrac1N G^{-1}$, so standard errors shrink as $1/\sqrt{N}$. The per-parameter
standard errors are the square roots of the diagonal of $G^{-1}$, obtained
analytically from the K-recursion Jacobian with no Monte Carlo; because they are
read from the diagonal of the \emph{full} inverse, each already accounts for
jointly estimating the remaining parameters rather than treating them as known.

When $G$ is rank-deficient (a gauge, as above), the bound is infinite
along $\ker G$: a parameter whose unit vector has a component in the null space
has unbounded variance and is flagged non-identifiable, with an infinite
standard error, while on the identifiable subspace the bound is the
Moore--Penrose pseudo-inverse of $G$ and its diagonal gives finite standard
errors. One Fisher object thus serves both roles at once: its null space is the
identifiability verdict, and its (pseudo-)inverse is the
reliability of the estimates.

\section{Numerical Experiments}
\label{sec:examples}
We exercise the whole framework on two examples: a linear Gaussian
state-space model (a first-order vector autoregression observed in noise), whose
chain structure admits validation against the classical Kalman recursions, and a
skip-connected extension of the same model, whose merging nodes make the parent
cross-covariances essential (\cref{sec:exp-skip}). The purpose is validation
rather than application benchmarking: each task carries either a classical
reference computation or a Monte-Carlo quantity against which the
covariance-chart result is checked. All numbers below are
produced by the reference implementation; where a closed form exists, we report
the agreement with it. The exact fitting settings and seeds are recorded in
\cref{app:exp}.

\subsection{Setup}
\label{sec:exp-setup}
\begin{figure}[t]
\centering
\includegraphics[width=0.92\linewidth]{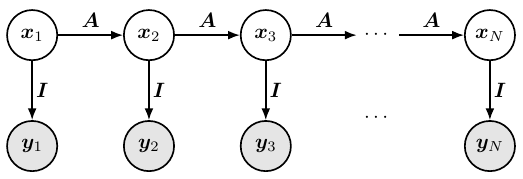}
\caption{The running example as a Bayesian network: a linear Gaussian
state-space model. Latent states $\bm{x}_1,\dots,\bm{x}_N$ (white) form a chain with the
transition $\bm{A}=\bm{S}\bm{T}$ tied across every step; each state emits a noisy
observation $\bm{y}_n=\bm{x}_n+\bm{w}_n$ (shaded). The first state has covariance $\bm{P}_0$;
process noise of covariance $\bm{Q}$ enters each subsequent state, and
observation noise of covariance $\bm{R}$ each $\bm{y}_n$
(\cref{eq:lds}).}
\label{fig:lds}
\end{figure}
The model has $d$-dimensional latent states $\bm{x}_1,\dots,\bm{x}_N$ and noisy
observations $\bm{y}_1,\dots,\bm{y}_N$,
\begin{equation}
\bm{x}_n=\bm{A}\bm{x}_{n-1}+\bm{u}_n,\qquad \bm{y}_n=\bm{x}_n+\bm{w}_n,
\label{eq:lds}
\end{equation}
with $\bm{x}_1\sim\mathcal{N}(\bm{0},\bm{P}_0)$, process noise
$\bm{u}_n\sim\mathcal{N}(\bm{0},\bm{Q})$, and observation noise
$\bm{w}_n\sim\mathcal{N}(\bm{0},\bm{R})$, all independent; following state-space
convention, these random vectors are written as boldface lowercase. As a
Bayesian network (\cref{fig:lds}) this
is a $2N$-node DAG: the state chain $\bm{x}_{n-1}\!\to\!\bm{x}_n$ carries the transition
$\bm{A}$, \emph{tied} across every step, and each $\bm{x}_n\!\to\!\bm{y}_n$ carries the
identity. The transition is itself structured as $\bm{A}=\bm{S}\bm{T}$ with
$\bm{S}$ a fixed, known matrix and $\bm{T}$ the learnable parameter, a concrete
instance of the tied/structured parametrization of \cref{rem:tied}. We take
$d=2$, $N=8$,
$\bm{S}=\left[\begin{smallmatrix}1&0.5\\0&1\end{smallmatrix}\right]$,
$\bm{T}=\left[\begin{smallmatrix}0.4&-0.6\\0.2&0.6\end{smallmatrix}\right]$ (so
$\bm{A}=\left[\begin{smallmatrix}0.5&-0.3\\0.2&0.6\end{smallmatrix}\right]$, with
spectral radius $0.6$), $\bm{P}_0=\bm{I}$, $\bm{Q}=0.1\,\bm{I}$, and
$\bm{R}=0.2\,\bm{I}$. This is the chain special case in which the K-recursion
reduces to the Kalman covariance recursion, which makes it an ideal validation
target rather than a new method. Throughout this section $M$ denotes the number
of i.i.d.\ trajectories, each a realization of the full $2N$-node network; $M$
plays the role of the sample count $N$ of
\crefrange{sec:estimation}{sec:identifiability}, since $N$ here is reserved for
the chain length.

\subsection{Inference and Conditional Independence}
\label{sec:exp-inf}
The framework assembles the joint covariance by the K-recursion and answers each
query by block algebra on it (\cref{sec:inference}). We check three facts against
the classical recursions.
\emph{Prior state covariance:} the marginal $\Cov[\bm{x}_n]$ read from the chart
matches the open-loop covariance recursion
$\bm{\Sigma}_n=\bm{A}\bm{\Sigma}_{n-1}\bm{A}^{\trans}+\bm{Q}$ (the Kalman
prediction step, applied with no measurement updates) to machine
precision (maximum block error $6.9\times10^{-18}$ over $n$), and
$\bm{\Sigma}_n$ approaches the fixed point of that recursion, the unique
solution $\bar{\bm{\Sigma}}$ of the discrete Lyapunov (Stein) equation
$\bar{\bm{\Sigma}}=\bm{A}\bar{\bm{\Sigma}}\bm{A}^{\trans}+\bm{Q}$, equivalently
$\bar{\bm{\Sigma}}=\sum_{k=0}^{\infty}\bm{A}^{k}\bm{Q}(\bm{A}^{\trans})^{k}$,
which exists because the spectral radius of $\bm{A}$ is $0.6<1$; at $n=N=8$
the gap is $\lVert\bm{\Sigma}_N-\bar{\bm{\Sigma}}\rVert=9.6\times10^{-4}$.
\emph{Smoothing:} conditioning the states on all observations,
$\Cov[\bm{x}_n\mid \bm{y}_{1:N}]$, is the Schur complement \eqref{eq:cond_cov}; it matches
the Rauch--Tung--Striebel smoother covariance to $1.5\times10^{-16}$.
Conditioning on the observations is therefore exactly Kalman smoothing, obtained
here as a special case of \cref{sec:conditioning}.
\emph{Conditional independence:} the Markov property surfaces as a vanishing CMI.
For a representative triple, $\MI(\bm{x}_{n-1};\bm{x}_{n+1}\mid \bm{x}_n)$ evaluates to exactly
zero in double precision, as \eqref{eq:ci} predicts, whereas conditioning off the
separating node leaves the two states dependent,
$\MI(\bm{x}_{n-1};\bm{x}_{n+1}\mid \bm{x}_{n+2})=0.156$ nats.

\subsection{Forward Sampling}
\label{sec:exp-sample}
Drawing trajectories by the forward pass of \cref{sec:sampling} and forming the
sample covariance reproduces the analytic $\bm{K}$: the relative error
$\lVert\hat{\bm{K}}-\bm{K}\rVert/\lVert\bm{K}\rVert$ is
$7.4\times10^{-2}$, $3.4\times10^{-2}$, $7.1\times10^{-3}$, $2.7\times10^{-3}$
at $M=10^{3},10^{4},10^{5},10^{6}$ trajectories, the $1/\sqrt{M}$ Monte-Carlo
rate.

\subsection{Estimation of the Tied Factor}
\label{sec:exp-est}
The data are $M=2\times10^{5}$ trajectories drawn from the true model of
\cref{sec:exp-setup} by the forward sampling of \cref{sec:exp-sample}; the
hidden states are then discarded, so the estimator sees only the observations
$\bm{y}_{1:N}$ of each trajectory. From these observations alone, and starting
from a random initialization, we estimate the shared factor $\bm{T}$ by
minimizing the negative marginal log-likelihood \eqref{eq:marglik} through the
differentiable chart, with $\bm{A}=\bm{S}\bm{T}$ tied across all $N-1$
transitions; the true $\bm{T}$ enters only the data generation and the final
error evaluation. The negative log-likelihood falls from $5.99$ to $0.99$ and
the estimate recovers the truth,
$\lVert\hat{\bm{T}}-\bm{T}\rVert=8.5\times10^{-4}$ (and
$\lVert\hat{\bm{A}}-\bm{A}\rVert=7.4\times10^{-4}$): a single $2\times2$ factor
is identified from the whole trajectory ensemble, with the gradient accumulated
over its shared occurrences by one AD sweep (\cref{rem:tied}).

\subsection{Identifiability and Reliability}
\label{sec:exp-crb}
\begin{figure}[t]
\centering
\includegraphics[width=\linewidth]{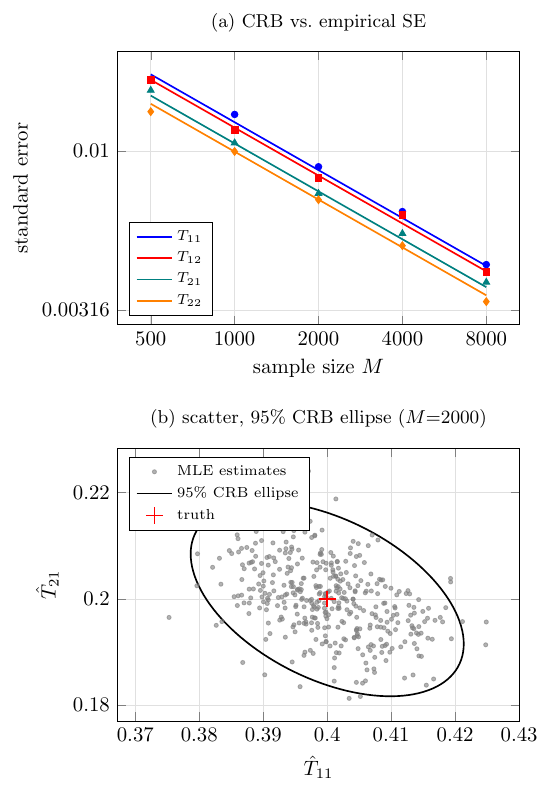}
\caption{Reliability of the tied factor $\bm{T}$ estimated from noisy
observations of the hidden states. (a)~Per-parameter standard error versus the
number of trajectories $M$ (log-log): the analytic Cram\'er--Rao bound
$\sqrt{\operatorname{diag}((MG)^{-1})}$ (lines) and the empirical scatter of the
maximum-likelihood estimate over $400$ trials (markers) agree and follow the
$1/\sqrt{M}$ slope. (b)~The $M=2000$ estimates $(\hat{T}_{11},\hat{T}_{21})$
(gray) with the analytic $95\%$ Cram\'er--Rao ellipse and the true value (red).}
\label{fig:crb}
\end{figure}

The Slepian--Bangs Fisher information over $\bm{T}$ (a $q=4$ parameter vector),
pulled back through $\bm{T}\mapsto\bm{K}_{\Obs\Obs}$ \eqref{eq:fisher}, has full
rank $4$ with eigenvalues $\{4.6,6.7,13.2,17.0\}$, so $\bm{T}$ is locally
identifiable from the observations (\cref{sec:gauge}), and its inverse sets the
reliability of the estimator. Over $400$ Monte-Carlo trials at each sample size
$M$, the empirical standard deviation of the maximum-likelihood estimate tracks
the analytic Cram\'er--Rao standard error
$\sqrt{\operatorname{diag}((MG)^{-1})}$ and follows the $1/\sqrt{M}$ law
(\cref{fig:crb}). At $M=2000$ the (analytic, empirical) per-parameter standard
errors are $(8.7,8.9)$, $(8.4,8.3)$, $(7.5,7.4)$, and $(7.0,7.0)$ for
$T_{11},T_{12},T_{21},T_{22}$ respectively, in units of $10^{-3}$: the analytic
bound predicts the observed estimator scatter to within the Monte-Carlo error.

\subsection{Beyond the Chain: Skip Connections}
\label{sec:exp-skip}
\begin{figure}[t]
\centering
\includegraphics[width=0.95\linewidth]{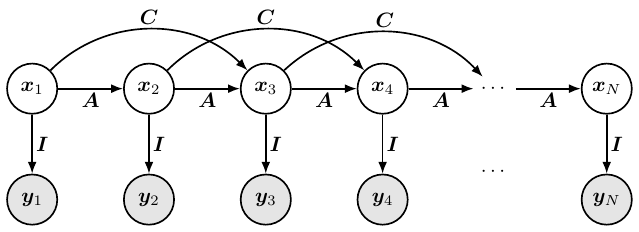}
\caption{The skip-connection extension of the running example
(\cref{sec:exp-skip}): the chain of \cref{fig:lds} plus tied skip edges
$\bm{x}_{n-2}\!\to\!\bm{x}_n$, each carrying the same matrix $\bm{C}$. Every
state with $n\geq3$ merges two correlated parents, so the parent
cross-covariances of \eqref{eq:krec} are essential and no first-order chain
recursion applies to the graph as given.}
\label{fig:skip}
\end{figure}
The second experiment takes the running example beyond the chain: a
\emph{tied skip connection} is added to \eqref{eq:lds},
\begin{equation}
\bm{x}_n=\bm{A}\bm{x}_{n-1}+\bm{C}\bm{x}_{n-2}+\bm{u}_n,\qquad n\geq3,
\label{eq:skip}
\end{equation}
with the $n=2$ step unchanged from \eqref{eq:lds}, $\bm{A}=\bm{S}\bm{T}$ and
all other settings as in \cref{sec:exp-setup}, and the known skip matrix
$\bm{C}=\left[\begin{smallmatrix}0.15&0\\0.05&0.1\end{smallmatrix}\right]$
shared across all skip edges (\cref{fig:skip}). Every state with $n\geq3$ now
merges two \emph{correlated} parents, so the parent cross-covariances in the
self-blocks of \eqref{eq:krec} become essential and the first-order recursions
of \cref{sec:exp-inf} no longer apply to the graph as given; the classical
route is a per-model reformulation, stacking $(\bm{x}_n,\bm{x}_{n-1})$ into a
first-order companion state. The framework needs no such reformulation: the
chart is simply re-run on the extended graph, and its marginals
$\Cov[\bm{x}_n]$ agree with the hand-augmented companion-form prediction
recursion to machine precision (maximum error $8.6\times10^{-17}$).

The information queries track the new structure. The first-order Markov CMI,
which vanished in \cref{sec:exp-inf}, is now
$\MI(\bm{x}_{n-1};\bm{x}_{n+1}\mid\bm{x}_n)=0.023$ nats, since the skip edge
keeps a path open around the conditioned node, while the second-order CMI
$\MI(\bm{x}_{n-1};\bm{x}_{n+2}\mid\bm{x}_n,\bm{x}_{n+1})$ is zero to double
precision, exactly as d-separation predicts on \cref{fig:skip}.

\begin{figure}[t]
\centering
\includegraphics[width=0.9\linewidth]{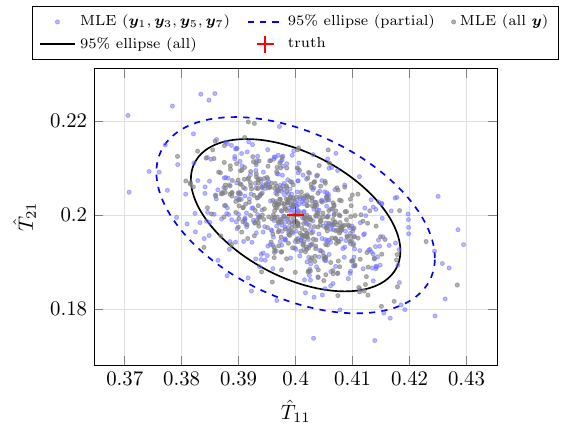}
\caption{Reliability of the tied factor $\bm{T}$ on the skip-connected model
of \cref{fig:skip}, for two observation patterns: all sensors observed (gray,
solid $95\%$ Cram\'er--Rao ellipse) and only the odd-indexed half
$\bm{y}_1,\bm{y}_3,\bm{y}_5,\bm{y}_7$ (blue, dashed ellipse). Each cloud shows
the $M=2000$ estimates $(\hat{T}_{11},\hat{T}_{21})$ over $400$
maximum-likelihood trials and the red cross the true value; each analytic
ellipse matches its own scatter, and the inflation of the dashed ellipse is
the information lost by halving the sensors.}
\label{fig:skipcrb}
\end{figure}

The estimation and reliability pipeline then runs unchanged. Repeating
\cref{sec:exp-est} verbatim on the extended graph, with the states hidden and
$M=2\times10^{5}$ trajectories, the tied factor is recovered from a random
initialization to $\lVert\hat{\bm{T}}-\bm{T}\rVert=5.5\times10^{-4}$; the
Fisher metric over $\bm{T}$ retains full rank $4$, and at $M=2000$ the
analytic Cram\'er--Rao standard errors $(7.5,7.3,6.6,6.2)$, in units of
$10^{-3}$, match the empirical scatter $(7.7,7.8,6.8,6.4)$ of $400$
maximum-likelihood trials. The same metric also prices observation patterns
with no further derivation: observing only the odd-indexed half of the
sensors (the observation nodes), $\bm{y}_1,\bm{y}_3,\bm{y}_5,\bm{y}_7$, keeps
$\bm{T}$ identifiable
(again full rank) but inflates the bound to $(10.0,9.6,8.5,7.9)$, matched by
the empirical $(10.1,10.1,8.6,7.7)$. \cref{fig:skipcrb} overlays the two
$95\%$ ellipses with their estimate clouds, the counterpart of
\cref{fig:crb}(b) on the extended graph: the cost of halving the sensors is
read off analytically, before any data are collected. Nothing in the code or
the derivations refers to
the topology: the same chart, likelihood, and Fisher pullback serve the chain
and its skip-connected extension alike.

\section{Conclusion}
\label{sec:conclusion}
We have developed inference, estimation, and identifiability for linear Gaussian
Bayesian networks as one differentiable covariance calculus on a single backend,
the K-recursion covariance chart $\Phi_{\mathcal{G}}:\param\mapsto\bm{K}$. Every
operation reduces to block algebra on the covariance it produces (block
selection, a Schur complement, a log-determinant, and a linear solve): marginals
and conditionals, mutual information and conditional-independence tests, forward
sampling, maximum-likelihood estimation under full and partial (hidden-node)
observation, and the
Slepian--Bangs Fisher information with the local identifiability and
Cram\'er--Rao reliability it induces. Because the chart is a smooth computation
graph, automatic differentiation supplies every gradient in one backward sweep,
over arbitrary vector-valued DAGs and arbitrary (including tied and structured)
parametrizations, with no per-query or per-topology derivation. On a linear
Gaussian state-space running example the framework reproduces the Kalman filter
and Rauch--Tung--Striebel smoother to machine precision, certifies the Markov
conditional independences as vanishing conditional mutual informations, recovers
a tied transition factor from noisy observations of hidden states, and yields an
analytic Cram\'er--Rao bound that matches the empirical scatter of the
maximum-likelihood estimator. On a skip-connected extension of the same
example, whose merging nodes take the graph beyond the chain, the identical
pipeline runs unchanged: the marginals match a hand-augmented companion-form
recursion, the conditional mutual information tracks d-separation, and the
tied factor is again recovered with Cram\'er--Rao-validated reliability.

The covariance map and its statistical ingredients are classical; the
contribution is their differentiable, unified formulation, realized in an
open-source reference implementation. The same backend drives
mutual-information \emph{optimization} in a companion paper~\cite{wadayama2026dag}.
Natural extensions include intervention and formal observation design on the
same chart, and a symbolic evaluation of the recursion for
machine-checked proofs of conditional independence and identifiability.

\appendix
The derivation of the K-recursion itself is given in the companion
paper~\cite{wadayama2026dag}. This appendix proves the two identities the paper
rests on, the log-determinant information formulas and the Slepian--Bangs
pullback, and records the settings behind the numbers of \cref{sec:examples}.

\subsection{Log-Determinant Identities \eqref{eq:mi}--\eqref{eq:ci}}
\label{app:logdet}
The differential entropy of a real Gaussian vector
$V_A\sim\mathcal{N}(\bm{0},\bm{K}_{AA})$ is
$h(V_A)=\tfrac12\log\det(2\pi e\,\bm{K}_{AA})$~\cite{coverthomas2006}. Because
the conditional covariance \eqref{eq:cond_cov} does not depend on the observed
value, the conditional entropy is
$h(V_A\mid V_B)=\tfrac12\log\det(2\pi e\,\bm{K}_{A\mid B})$, and
$\MI(V_A;V_B)=h(V_A)-h(V_A\mid V_B)$ gives \eqref{eq:mi}, the $2\pi e$ factors
canceling. Conditioning every entropy on $V_C$ gives \eqref{eq:cmi}.

For \eqref{eq:ci}, condition the pair $(V_A,V_B)$ on $V_C$: the joint
conditional covariance has diagonal blocks $\bm{K}_{A\mid C},\bm{K}_{B\mid C}$
and off-diagonal block $\bm{K}_{AB\mid C}$, and conditioning further on $V_B$
gives the nested Schur complement
$\bm{K}_{A\mid BC}
 =\bm{K}_{A\mid C}-\bm{K}_{AB\mid C}\bm{K}_{B\mid C}^{-1}\bm{K}_{BA\mid C}$.
Substituting into \eqref{eq:cmi},
\begin{equation}
\MI(V_A;V_B\mid V_C)=-\tfrac12\log\det\!\big(\bm{I}-\bm{R}\bm{R}^{\trans}\big),
\label{eq:cmi_cc}
\end{equation}
with $\bm{R}=\bm{K}_{A\mid C}^{-1/2}\bm{K}_{AB\mid C}\bm{K}_{B\mid C}^{-1/2}$.
The singular values $\rho_i$ of $\bm{R}$ are the canonical correlations of
$V_A$ and $V_B$ given $V_C$ and lie in $[0,1)$, so
$\MI(V_A;V_B\mid V_C)=-\tfrac12\sum_i\log(1-\rho_i^2)\geq0$, with equality if
and only if every $\rho_i=0$, that is, $\bm{K}_{AB\mid C}=\bm{0}$. This proves
\eqref{eq:ci}, and nonnegativity along the way. In the circular-complex case
the entropy is $\log\det(\pi e\,\bm{K})$, which is where the factor $\tfrac12$
of \eqref{eq:mi}--\eqref{eq:cmi} becomes $1$ (\cref{rem:complex}).

\subsection{The Slepian--Bangs Pullback \eqref{eq:fisher}}
\label{app:sb}
Write $\bm{K}\equiv\bm{K}_{\Obs\Obs}(\param)\succ\bm{0}$ (\cref{sec:psd}) and
$\partial_a\equiv\partial/\partial\eta_a$. The log-density of one observation
$v\sim\mathcal{N}(\bm{0},\bm{K})$ is
$\log p(v)=-\tfrac12\log\det(2\pi\bm{K})-\tfrac12 v^{\trans}\bm{K}^{-1}v$.
The matrix identities
$\partial_a\log\det\bm{K}=\tr(\bm{K}^{-1}\partial_a\bm{K})$ and
$\partial_a\bm{K}^{-1}=-\bm{K}^{-1}(\partial_a\bm{K})\bm{K}^{-1}$ give the
score in centered quadratic-form shape,
\begin{equation}
s_a(v)=\tfrac12\big(v^{\trans}\bm{W}_a v-\tr(\bm{W}_a\bm{K})\big),
\quad
\bm{W}_a=\bm{K}^{-1}(\partial_a\bm{K})\bm{K}^{-1},
\label{eq:score}
\end{equation}
using $\E[v^{\trans}\bm{W}_a v]=\tr(\bm{W}_a\bm{K})$. Each $\bm{W}_a$ is
symmetric, and for a zero-mean Gaussian the covariance of two quadratic forms
is $\Cov(v^{\trans}\bm{W}_a v,\,v^{\trans}\bm{W}_b v)
=2\tr(\bm{W}_a\bm{K}\bm{W}_b\bm{K})$ (Isserlis' theorem). Hence
$G_{ab}=\E[s_a s_b]=\tfrac12\tr(\bm{W}_a\bm{K}\bm{W}_b\bm{K})$, and expanding
$\bm{W}_a,\bm{W}_b$,
\[
G_{ab}=\tfrac12\tr\!\big(\bm{K}^{-1}\partial_a\bm{K}\,\bm{K}^{-1}\partial_b\bm{K}\big),
\]
which is \eqref{eq:fisher}. The parameters enter only through
$\bm{K}_{\Obs\Obs}$, so this is the Fisher information of the Gaussian family
pulled back along $\param\mapsto\bm{K}_{\Obs\Obs}$ by the chain rule, and the
derivatives $\partial_a\bm{K}_{\Obs\Obs}$ are exactly the AD Jacobian of the
K-recursion. The same computation with the circular-complex density and the
complex form of Isserlis' theorem yields the factor $1$ of \cref{rem:complex}.

\subsection{Experimental Details}
\label{app:exp}
All numbers and figure data of \cref{sec:examples} are produced by two scripts
of the reference implementation, one per example, with fixed random seeds, so
every reported digit is reproducible by one command each. Four settings deserve
explicit mention. (i)~In \cref{sec:exp-est} the tied factor is estimated by minimizing
the negative marginal log-likelihood \eqref{eq:marglik} with Adam (learning rate $0.02$,
$1500$ iterations) from a random initialization; the reported objective values
are per-sample negative log-likelihoods up to the usual constant.
(ii)~In \cref{sec:exp-crb} each of the $400$ trials per sample size draws $M$
fresh trajectories, forms the observed sample covariance, and computes the
maximum-likelihood estimate by Newton's method on \eqref{eq:marglik} ($10$
iterations, gradient and Hessian by AD), initialized at the true parameter so
that every trial converges to the consistent root of the likelihood equation,
the standard practice when checking Cram\'er--Rao attainment.
(iii)~The Fisher matrix $G$ of \eqref{eq:fisher} is evaluated at the true
parameter from the AD Jacobian, and the ellipse of \cref{fig:crb}(b) is the
$95\%$ level set of the Gaussian whose covariance is the corresponding
$2\times2$ sub-block of $(MG)^{-1}$ at $M=2000$.
(iv)~In \cref{sec:exp-skip} the tied factor is estimated with the same
settings as~(i), and the Monte-Carlo validation uses the same Newton procedure
as~(ii), with $400$ trials of $M=2000$ draws for each of the two observation
patterns; the ellipses of \cref{fig:skipcrb} are constructed as in~(iii).

\section*{Acknowledgment}
This work was supported by JST, CRONOS, Japan Grant Number JPMJCS25N5.


\end{document}